\documentclass[a4paper,fleqn,usenatbib,useAMS]{mnras}

\usepackage{graphicx}	
\usepackage{amsmath}	
\usepackage{amssymb}	
\usepackage{multicol}        
\usepackage{bm}		
\usepackage{pdflscape}	

\usepackage[T1]{fontenc}
\usepackage{ae,aecompl}

\usepackage{newtxtext,newtxmath}
\usepackage{tikz}

\title[Variable quiescent state for SAX J1750.8-2900]{Variable quiescent state for the neutron-star X-ray transient SAX J1750.8-2900: not such a hot neutron star after all?}

\author[A.S. Parikh and R. Wijnands]{A.S. Parikh$^{1}$\thanks{Contact e-mail: \href{mailto:a.s.parikh@uva.nl}{a.s.parikh@uva.nl}} and
R. Wijnands$^{1}$
\\
$^{1}$Anton Pannekoek Institute for Astronomy, University of Amsterdam, Postbus 94249, 1090 GE Amsterdam, The Netherlands}

\pubyear{2016}

\begin{document}
\label{firstpage}
\pagerange{\pageref{firstpage}--\pageref{lastpage}}
\maketitle

\begin{abstract}
We monitored the neutron star low-mass X-ray binary SAX J1750.8$-$2900 after the end of its 2015/2016 outburst using the X-ray Telescope (XRT) aboard {\it Swift} to detect possible post-outburst `rebrightenings', similar to those seen after its 2008 outburst. We did not detect any rebrightening behaviour, suggesting that the physical mechanism behind the rebrightening events is not always active after each outburst of the source. Any model attempting to explain these rebrightenings should thus be able to reproduce the different outburst profiles of the source at different times. Surprisingly, our {\it Swift}/XRT observations were unable to detect the source, contrary to previous {\it Swift}/XRT observations in quiescence. We determined a temperature upper limit of $\leq$ 106 eV, much colder than the post 2008 outburst value of $\sim$ 145 eV. We also report on an archival {\it Chandra} observation of the source after its 2011 outburst and found a temperature of $\sim$ 126 eV. These different temperatures, including the non-detection very close after the end of the 2015/2016 outburst, are difficult to explain in any model assuming we observe the cooling emission from a neutron star core or an accretion-heated crust. We discuss our observations in the context of a change in envelope (the outer $\sim$ 100 m of the crust) composition and (possibly in combination with) a cooling crust. Both hypotheses cannot explain our results unless potentially unrealistic assumptions are made. Irrespective of what causes the temperature variability, it is clear that the neutron star in SAX J1750.8$-$2900 may not be as hot as previously assumed.
\end{abstract}





\section{Introduction}
Low-mass X-ray binaries (LMXBs) consist of a neutron star or black hole having a sub-solar donor star. This companion star facilitates accretion onto the primary by overflowing its Roche lobe. Most systems are not accreting persistently but only during outbursts which have typical X-ray luminosities of $L_\text{X} \sim 10^{35}$ -- $10^{39}$ erg s$^{-1}$. The outburst episodes are separated by long periods of quiescence in which the X-ray luminosity is very low ($L_\text{X} \sim 10^{30}$ -- $10^{34}$ erg s$^{-1}$). During the outbursts, the accretion rate onto the compact object varies strongly. Commonly, these outbursts are explained by the disk instability model \citep[see][for a review]{lasota2001disc}, although many uncertainties remain. Several systems show post-outburst `rebrightenings'\footnote{`Rebrightenings' have also been referred to as `mini outbursts', `echo outbursts', and `reflares' in the literature \citep{osaki2001repetitive,csak2005spectroscopic,patruno2016reflares}} that may appear multiple times at low luminosities (10$^{32}$--10$^{36}$ erg s$^{-1}$), lasting for days to tens of days. Among these systems are both neutron star and black hole transients. Furthermore, they are also observed in transiently accreting white dwarfs (during so-called dwarf novae), suggesting that the production mechanism is related to the general behaviour of accretion flows and not connected to a specific type of accretor \citep[see][for a discussion]{patruno2009saxj1808}. These rebrightenings cannot be explained within the disk instability model in a straightforward manner \citep[e.g.,][]{lasota2001disc,patruno2009saxj1808,kotko2012models}.

For LMXBs having neutron stars (NSs), the accreted material during an outburst compresses the surface and results in crust heating by processes such as electron capture, pycnonuclear reactions, and neutron emission \citep[e.g.,][]{haensel1990non,haensel2008models,steiner2012deep}. This results in the crust being heated out of equilibrium with the core. Once in quiescence this heated crust cools to reinstate thermal equilibrium with the core. Thus, any heating and cooling observed in this scenario is indicative of a change in the crust temperature. The core does not show any significant temperature evolution over these time scales. Monitoring the cooling evolution of an accretion-heated crust can help us understand the physics of the neutron star crust \citep[e.g.,][]{rutledge2002crustal,brown2009mapping,page2013forecasting}.  
So far ten NS LMXBs that potentially exhibit crust cooling have been studied (see e.g., \citeauthor{wijnands2001chandra} \citeyear{wijnands2001chandra}, \citeauthor{degenaar2017cold} \citeyear{degenaar2017cold}, \citeauthor{parikh2017potential} \citeyear{parikh2017potential}; see \citeauthor{wijnands2017review} \citeyear{wijnands2017review} for a review).

SAX J1750.8$-$2900 (hereafter SAX J1750) was first detected by {\it BeppoSAX} in 1997 \citep{natalucci1999new}. The source showed several type-I thermonuclear bursts demonstrating the primary to be a NS. Some of the type-I bursts showed photospheric radius expansions from which an upper limit on the distance was calculated to be 6.79 $\pm$ 0.14 kpc \citep[assuming H-poor burning;][]{kaaret2002discovery,galloway2008thermonuclear}. \citet{kaaret2002discovery} observed nearly coherent oscillations from the source during its type-I bursts, revealing the spin of the neutron star to be $\sim$ 601 Hz. Since its first detection, four more outbursts have been observed -- the longer outbursts (duration of $\sim$ 400 d) in 2001, 2008, and 2015/2016 \citep[][Figure \ref{fig_lc_all}]{kaaret2002discovery,markwardt2008rxte,sanchez2015integral}; and a shorter outburst ($\sim$ 30 d) in 2011 \citep{fiocchi2011integral,natalucci2011swift}.  

The source has been studied in quiescence after its 2008 and 2011 outbursts. After its 2008 outburst, SAX J1750 did not go straight into quiescence but it showed a period of rebrightenings that lasted for $\sim$ 200 d (\citeauthor{lowell2012xmm} \citeyear{lowell2012xmm}, see also \citeauthor{allen2015spectral} \citeyear{allen2015spectral}). The source was found truly in quiescence during an  {\it XMM-Newton} observation taken $\sim$ 400 d after the full outburst with a relatively high surface temperature of $\sim$ 148 eV, resulting in the speculation that it might host the hottest known NS in a LMXB\footnote{It was assumed that the crust and core were in equilibrium at the time of this {\it XMM-Newton} observation.} \citep{lowell2012xmm}. \citet{wijnands2013low} studied the source $\sim$ 350 d after the 2011 outburst when it showed a brief accretion flare in its quiescent state. The pre-flare quiescent level they found was in agreement with that found after the 2008 outburst by \citet{lowell2012xmm}. 

\section{Observations and Data Analysis}
We monitored SAX J1750, using the X-ray telescope \citep[XRT; ][]{burrows2005swift} on board {\it Swift}, after the end of its 2015/2016 outburst to investigate if the source also exhibited rebrightenings after this recent outburst. Surprisingly, we found that the source was not detected at all, and so no rebrightenings were observed. The upper limits (on the luminosity and inferred surface temperature, see Section \ref{sec_swift_analysis}) determined for these observations was below those determined previously by {\it XMM-Newton} and earlier {\it Swift}/XRT observations when the source was in quiescence \citep{lowell2012xmm,wijnands2013low}. To compare our new results with these earlier findings and to ensure uniform analysis we reanalysed the archival quiescent {\it XMM-Newton} and {\it Swift}/XRT data. In addition, to study further variability in quiescence we searched the {\it XMM-Newton}, {\it Swift}/XRT, and {\it Chandra} archives for additional observations of the source. We found one extra {\it Chandra} observation (performed in 2013) when the source was in quiescence. We also analyse this observation in our paper. We use the {\it Swift}/Burst Alert Telescope \citep[BAT;][]{barthelmy2005burst} and the {\it Rossi X-ray Timing Explorer}/Proportional Counting Array \citep[{\it RXTE}/PCA;][]{jahoda2006calibration} to track the luminosity evolution of the source over time.

\subsection{{\textit{\textbf{Swift}}}/BAT and \textit{\textbf{RXTE}}/PCA}

We use the data from the {\it Swift}/BAT and the {\it RXTE}/PCA instruments to track the long-term outburst variability of SAX J1750 in the 15 -- 50 keV and 2 -- 60 keV range, respectively. We downloaded the {\it Swift}/BAT data for SAX J1750 from the BAT online archive\footnote{http://swift.gsfc.nasa.gov/results/transients/} \citep{krimm2013swift}. The {\it RXTE}/PCA data was taken from the {\it RXTE} Galactic Center Observation archive, compiled by Craig Markwardt\footnote{https://asd.gsfc.nasa.gov/Craig.Markwardt//galscan/html/SAXJ1750.8-2900.html} \citep{swank2001populations}. The {\it Swift}/BAT data were rebinned with a bin size of 10 d.

\subsection{\textit{XMM-Newton}}
\label{sect_dat_an_xmm}

\begin{figure}
\centering
\begin{tikzpicture}
	\node[anchor=south west,inner sep=0] at (0,0) {\includegraphics[scale=2.05]{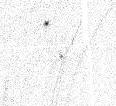}};
    \draw[black,thick,dashed](4.43,3.62) circle (0.34cm);
    \draw (8.05,0.5) -- (7.234,0.5) ;
     \draw (8.05,0.6) -- (8.05,0.4);
         \draw (7.234,0.6) -- (7.234,0.4);
 \node at (7.6,0.25) {1 arcminute};
\end{tikzpicture}
\caption{The image of the field near SAX J1750.8$-$2900 (indicated by the dashed circle of radius 25 arcsec) as obtained using the EPIC-MOS1 instrument on board \textit{XMM-Newton}. The source located to the upper left is 2RXP J175029.3$-$285954. The stray light contamination seen in the figure is caused by another nearby bright source (outside the field of view).}
\label{imag_src_mos1}
\end{figure}
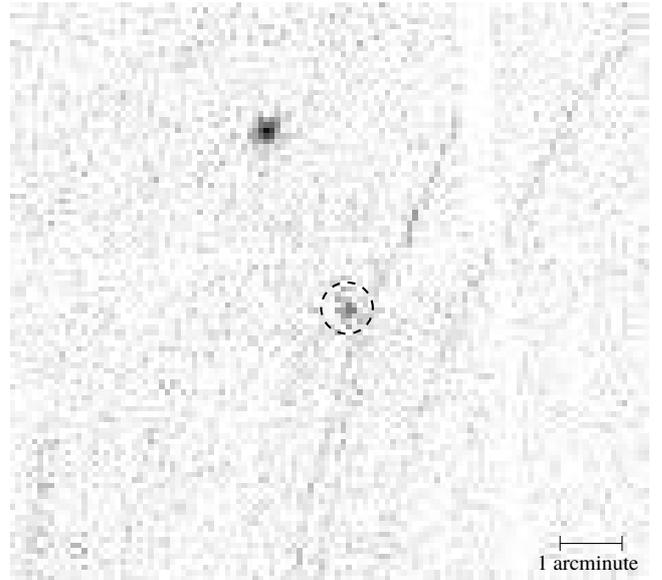

{\it XMM-Newton} was used to observe SAX J1750 on 2010 April 7 (observation ID [ObsID]: 0603850201) using all three EPIC instruments -- MOS1, MOS2, and pn, operated in window mode. The data were downloaded from the {\it XMM-Newton} archive\footnote{http://nxsa.esac.esa.int/nxsa-web/\#search}. We did not use the RGS and OM data because the source was too faint to be studied using these instruments. The $\sim$ 34 ksec raw data were reprocessed using the Science Analysis System (\texttt{SAS}; version 14.0) with \texttt{emproc} and \texttt{epproc}. The light curve of the data were checked for background flaring at high energies, $>$ 10 keV for the MOS and between 10 -- 12 keV for the pn. Data with count rates (in these energy ranges) above 0.2 counts s$^{-1}$ and 0.5 counts s$^{-1}$ were removed from the MOS and pn data, respectively. The final exposure time of the MOS1, MOS2, and pn was 19.6 ksec, 19.6 ksec, and 16.2 ksec, respectively. Circular source extraction regions of radius 20 arcsec and 25 arcsec were used for the MOS and pn, respectively. All instruments suffered from contamination by stray light due to a nearby bright source outside the field of view. This can be seen in Figure \ref{imag_src_mos1} which shows the field of view around SAX J1750. The figure also shows another nearby source in the field of view -- 2RXP J175029.3$-$285954 \citep{jonker2011galactic}. A stripe of stray light was seen crossing the source position, as observed by all three cameras. The background regions were circular regions of radius 25 arcsec for the MOS and pn detectors, placed on the same CCD as that of the source. To correct for the stray light contamination, background regions were selected such that this stripe was included in them. The redistribution matrix files and ancillary response functions were generated using \texttt{rmfgen} and \texttt{arfgen}. The spectra were grouped to have a minimum of 10 photons per bin using \texttt{specgroup}.

\subsection{\textit{\textbf{Swift}}/XRT}
\label{sect_obs_xrt}
All the {\it Swift}/XRT observations of SAX J1750, up to October 2016, were used to create a light curve. The data were downloaded from the HEASARC archive\footnote{http://heasarc.gsfc.nasa.gov/cgi-bin/W3Browse/swift.pl}. The raw data were processed with \texttt{xrtpipeline} using \texttt{HEASOFT} (version 6.17). The \texttt{XS\textsc{elect}} (version 2.4c) program was used for obtaining the light curve using a circular source region with a radius of 50 arcsec and an annulus having an inner radius of 250 arcsec and an outer radius of 350 arcsec as the background region. Neither of the regions include the nearby source 2RXP J175029.3$-$285954. No type-I thermonuclear bursts were detected in any of the XRT data. None of the Window Timing (WT) mode data were piled-up but some Photon Counting (PC) mode data (7 observations) were piled-up. This was corrected for by removing the data located in the inner source region (the exact radius of the exclusion region was calculated according to the online thread\footnote{http://www.swift.ac.uk/analysis/xrt/pileup.php}). 

Out of all the XRT observations, we can identify two sets of quiescent observations -- one after the 2011 outburst and one after the 2015/2016 outburst. All these data were collected using the PC mode. We have stacked the data after each outburst to obtain constraints on the observed effective surface temperature and luminosity. Interval 1 (post 2011 outburst; see Table 1 of \citeauthor{allen2015spectral} \citeyear{allen2015spectral}) ranges from 2012 Feb 14 (ObsID: 0031174024) to 2012 Mar 6 (ObsID: 0031174028). A flare was detected on 2012 Mar 17 and 20 \citep[ObsID: 0031174030 and ObsID: 0031174031;][]{wijnands2013low} and these data have been excluded in the determination of the quiescent level. Studying the source images of the observations taken after the flare reveals that for all but one of these observations the source was near the edge of the CCD. The observation that was centred on the source (Obs ID: 0031174032) had an exposure of 1 ksec and only provided a non-constraining upper limit. Thus, we also did not use the post-flare observations to determine the quiescent level of SAX J1750. The observations that were obtained after the 2015/2016 outburst are combined as Interval 2 (see Table \ref{tab_log_obs}). The observations making up each interval were stacked to a create two combined event files, one for each of the two intervals. 

The source was not detected in Interval 2 and a spectrum could not be extracted. Instead we simulated a spectra to obtain upper limits (see Section \ref{sec_swift_analysis}). The spectrum for Interval 1 was extracted using \texttt{XS\textsc{elect}} with a circular region of radius 25 arcsec for the source. For the background, regions comprising of three circular regions having a radius of 25 arcsec each were used. The background regions were chosen to ensure that the nearby source 2RXP J175029.3$-$285954 (Fig. \ref{imag_src_mos1}) was not included. The {\it Swift}/XRT data do not show any evidence of contamination from the other nearby bright source (that lies outside the field of view) which results in stray light in the {\it XMM-Newton} data (see Section \ref{sect_dat_an_xmm}). The sizes of the source and background regions used here are smaller than those used to create the light curve (see first paragraph of this section) as all the quiescent data are observed in the PC mode and do not have many photons ($\sim$ 0 -- 20). The stacked exposure map was created using \texttt{ximage}. The resulting spectrum were grouped to have a minimum of 2 counts per bin using \texttt{grppha}. The ancillary response files were generated using \texttt{xrtmkarf} and the appropriate \texttt{rmf} files were used\footnote{Interval 1 : swxpc0to12s6$\_$20110101v014.rmf\\  Interval 2 : swxpc0to12s6$\_$20130101v014.rmf}.  

\subsection{\textit{Chandra}}
SAX J1750 was observed using {\it Chandra} on 2013 April 29 for 25 ksec in the faint mode with the ACIS-S array. We downloaded the data (obsID: 14651) from the {\it Chandra} Data Archive\footnote{http://cda.harvard.edu/chaser/} and used \texttt{CIAO} (version 4.8) to analyse it. The {\it Chandra} data were examined for possible background flaring by studying the light curve from the background region (all parts of the detector except the source region). No background flaring was present for the {\it Chandra} observation of SAX J1750, so we used the entire exposure time of 25 ksec for our spectral analysis. 

A circular source region of radius 2 arcsec was used  for light curve and spectrum extraction. A background consisting of four circular regions, each having a radius of 10 arcsec, was used. The background regions were placed on the same CCD as that of the source and avoid the bright source in the field of view (2RXP J175029.3$-$285954; see Fig. \ref{imag_src_mos1}). There was no evidence of contamination by stray light from the other nearby source outside the field of view (see Section \ref{sect_dat_an_xmm}). The spectrum was extracted using \texttt{specextract} and rebinned using \texttt{grppha} to have at least 5 photons per bin. We used the auxiliary response file created by \texttt{specextract} for which the point-source aperture correction has been applied.

\section{Results}

\begin{figure}
\centering
\includegraphics[scale=0.58]{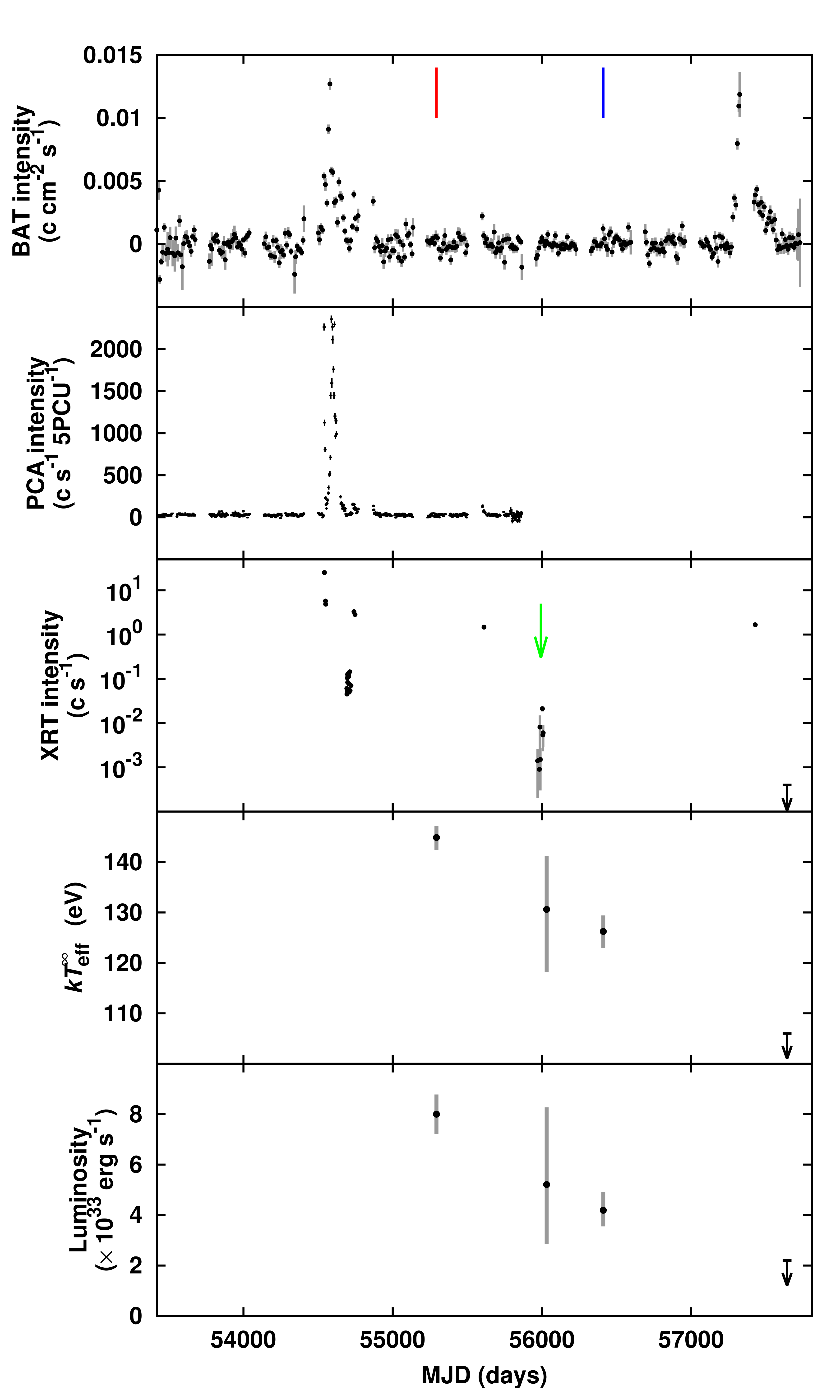}
\caption{The top three panels show the light curve of SAX J1750 obtained using {\it Swift}/BAT (15 -- 50 keV, binsize = 10 d), {\it RXTE}/PCA (2 -- 60 keV, binsize = 1 d), and {\it Swift}/XRT (0.5 -- 10 keV), respectively. The vertical red and blue lines in the top panel indicate the times of the {\it XMM-Newton} observation and the {\it Chandra} observation, respectively. The green arrow in the XRT light curve (middle panel) indicates the time when the flare reported by \citet{wijnands2013low} occurred. The bottom two panels show the observed effective temperature and luminosity evolution of the source in quiescence.}
\label{fig_lc_all}
\end{figure}

\begin{figure}
\centering
\includegraphics[scale=0.59]{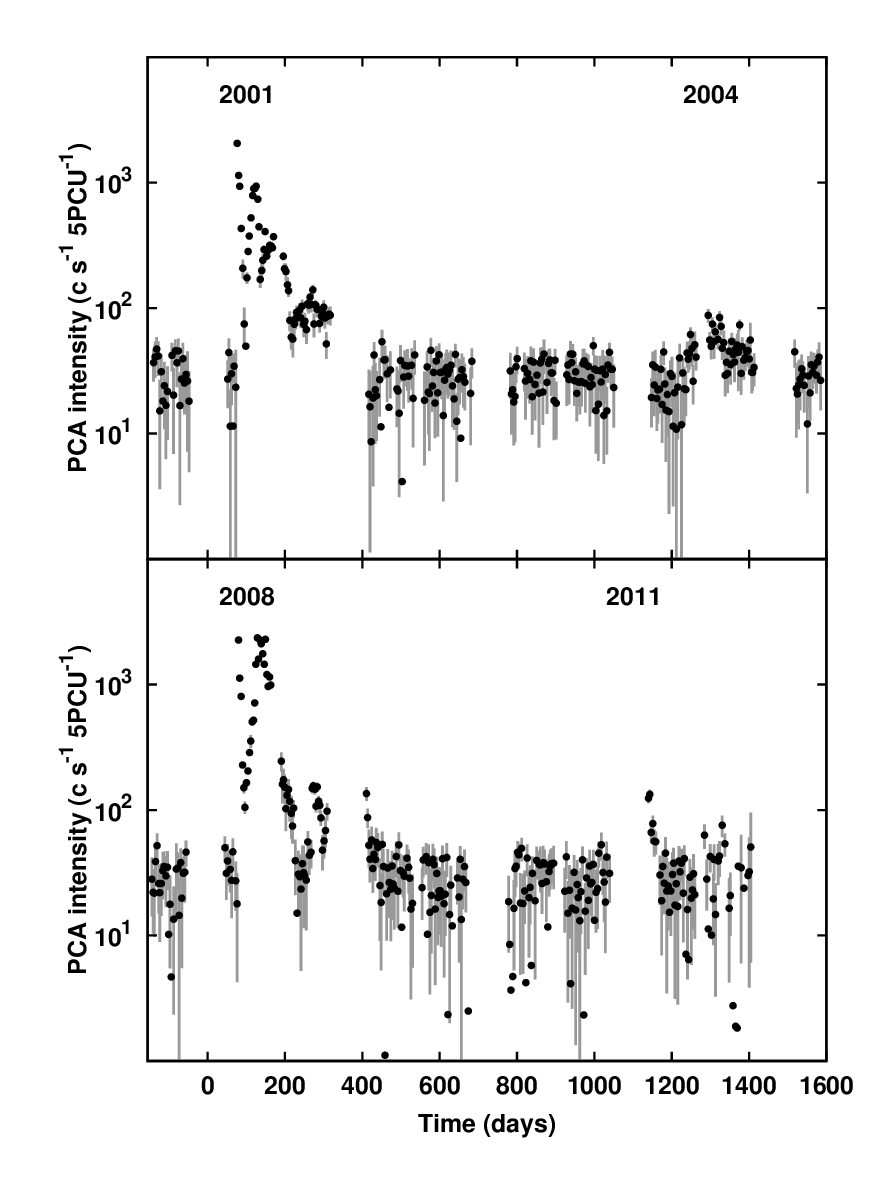}
\caption{The \textit{RXTE}/PCA light curves (2 -- 60 keV, binsize = 1 d) of the 2001 and 2004 outbursts (top panel), and the 2008 and 2011 outbursts (bottom panel) of SAX J1750. The zero point for the time axis of the top panel is MJD 51940 and for the bottom panel MJD 54460.}
\label{fig_2001_2008_lc}
\end{figure}

\begin{figure}
\centering
\includegraphics[scale=0.59]{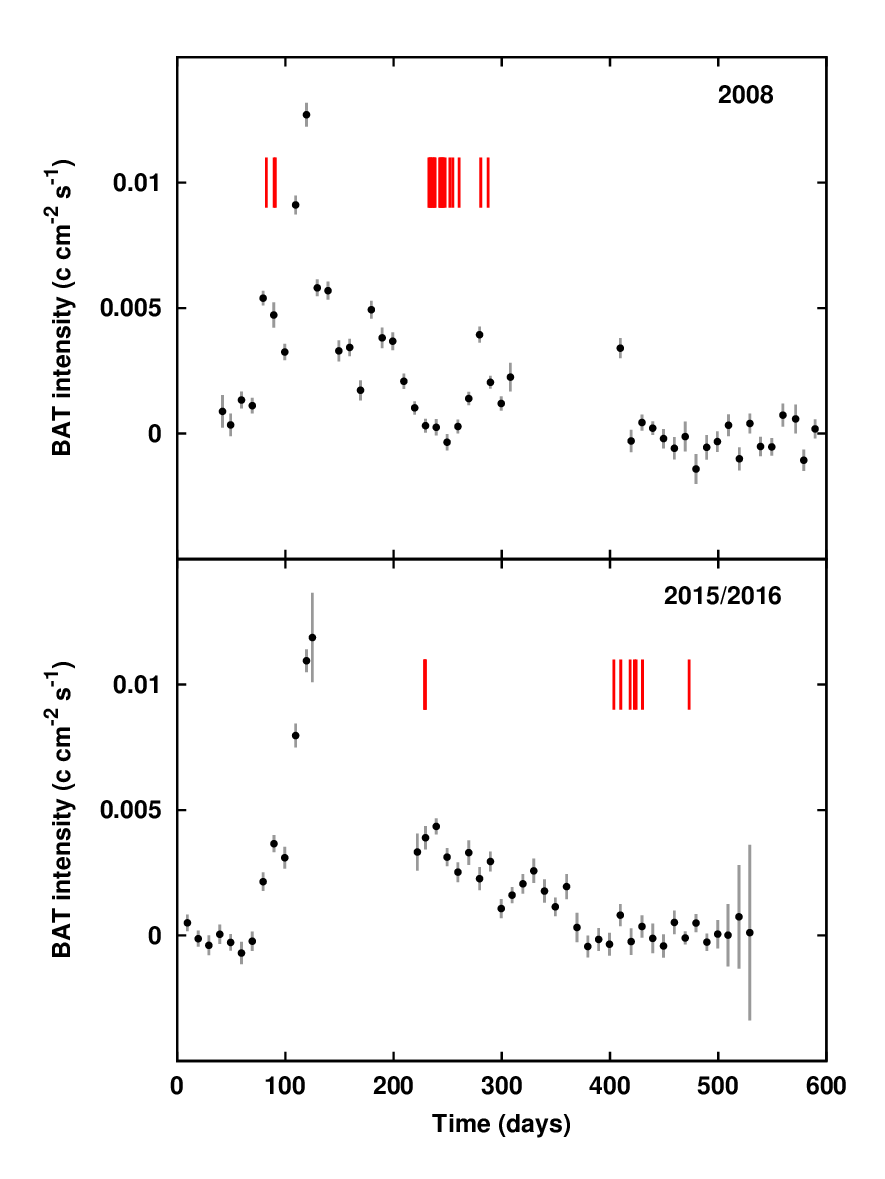}
\caption{A zoom in of the \textit{Swift}/BAT light curve (15 -- 50 keV, binsize = 10 d) of SAX J1750 of its 2008 (top panel) and 2015/2016 (bottom panel) outbursts. The zero point for the time axis for the 2008 outburst is MJD 54460 and for the 2015/2016 outburst MJD 57200. The red vertical lines in both panels indicate the times of the \textit{Swift}/XRT observations.}
\label{fig_2008_2015_lc}
\end{figure}

\subsection{Outburst Properties}

The {\it Swift}/BAT and {\it RXTE}/PCA light curves of SAX J1750 (Fig. \ref{fig_lc_all}, first and second panel) show the 2008, 2011, and 2015/2016 outbursts. 
A close up of the 2008 outburst (including the rebrightenings) is also shown in Figure \ref{fig_2001_2008_lc}, lower panel (PCA data) and Figure \ref{fig_2008_2015_lc}, upper panel (BAT data). 
In Figure \ref{fig_2001_2008_lc}, the 2011 outburst is visible as a slight rise in count rate around MJD 55700 when the source became visible again after the Sun time constraint window ended. Likely, most of the 2011 outburst occurred at the time of the Sun constraint window. The total observed outburst duration, post the Sun constraint window, was $\sim$ 20 d which indicates a lower limit on the length of the outburst. The upper limit on the length of the 2011 outburst is $\sim$ 110 d (the $\sim$ 20 d of the observed outburst plus the $\sim$ 90 d the source was in the Sun constraint window). The brightest reported luminosity during this outburst was $L_\text{X} \sim 10^{36}$ erg s$^{-1}$ \citep{natalucci2011swift}. 

\citet{sanchez2015integral} reported the initial detection of the 2015/2016 outburst of SAX J1750 on MJD 57278. 
The PCA also observed the 2001 outburst of the source and the light curve of this outburst is shown in Figure \ref{fig_2001_2008_lc} (upper panel).
 In this figure a smaller outburst $\sim$ 1000 d after the end of the 2001 outburst can be seen. This outburst has not yet been reported. We shall refer to it as the 2004 outburst. As can be seen from Figure \ref{fig_2001_2008_lc} and Figure \ref{fig_2008_2015_lc} the 2001, 2008, and 2015/2016 outbursts seem to have ended in $\sim$ 300 -- 400 d.


The background subtracted and pile-up corrected {\it Swift}/XRT light curve of all observations of SAX J1750 is shown in Figure \ref{fig_lc_all} (middle panel). The 2008 outburst has been previously studied by \citet{lowell2012xmm} and \citet{allen2015spectral}. The 2015/2016 outburst of SAX J1750 was not well covered with only one observation. During our {\it Swift}/XRT monitoring campaign after the 2015/2016 outburst, we obtained data using six pointings in August 2016 for a total of  $\sim$ 5 ksec (see Table \ref{tab_log_obs} for the log of the observations). Surprisingly, we did not detect the source, indicating that during our observations the source did not show any rebrightening events. To confirm the source would not exhibit such a rebrightening event in a later phase after the outburst and to get better upper limits in case the source was once again not detected, we obtained another $\sim$ 4 ksec XRT observation in October 2016. During this observation we, once again, did not detect the source. In Section \ref{sec_swift_analysis}, we discuss these observations further when we  determine luminosity and temperature  upper limits.

\begin{table}
\centering
\caption{Log of the {\it Swift}/XRT observations of SAX J1750 after the end of its 2015/2016 outburst, forming Interval 2. The upper limits on the count rate are determined for a 95$\%$ confidence level using the prescription by \citet{gehrels1986confidence}.}
\label{tab_log_obs}
\begin{tabular}{rccc}
\hline
Date & Observation & Exposure & Count Rate \tabularnewline
& ID & Time & (0.5 -- 10 keV) \tabularnewline
& & (ksec)& (10$^{-3}$ c s$^{-1}$) \tabularnewline
 \hline
  2016/08/03 &00031174033 & 0.9 & $<$3.2          \tabularnewline
09 &00031174035 & 0.3 & $<$1.2         \tabularnewline
18 &00031174037 & 1.0 & $<$3.1          \tabularnewline
22 &00031174038 & 0.9 & $<$3.5          \tabularnewline
24 &00031174039 & 1.3 & $<$2.4          \tabularnewline
30 &00031174040 & 0.5 & $<$6.1          \tabularnewline
2016/08& Total& 4.8 & $<$0.6\tabularnewline
  2016/10/12 &00031174041 & 3.7 & $<$0.8        \tabularnewline
 & Total & 8.5 & $<$0.4 \tabularnewline

\hline
\end{tabular}
\end{table}

\begin{table}
\centering
\caption{Obtained temperatures and luminosities of SAX J1750$^{\text{a}}$.}
\label{tab_temp_and_lumin}
\begin{tabular}{p{0.8cm}p{1.6cm}p{1.3cm}p{1cm}p{1.6cm}}
\hline
MJD & Instrument & Observation & $kT_{\text{eff}}^{\infty}$ & Luminosity$^\text{b}$ \tabularnewline
(days) & & ID & (eV) & (10$^{33}$ erg s$^{-1}$;\tabularnewline
& & & &  0.5 -- 10 keV) \tabularnewline
\hline
55293.5 & {\it XMM-Newton} & 0603850201 & 144.9$_{-2.5}^{+2.3}$ &8.0 $\pm$ 0.8\tabularnewline
56031.7 & {\it Swift}/XRT & Interval 1$^\text{c}$ & 130.6$_{-12.5}^{+10.6}$ & $5.2_{-2.4}^{+3.1}$\tabularnewline
56411.8 & {\it Chandra} & 14651 & 126.2$_{-3.3}^{+3.2}$ & 4.2 $\pm$ 0.7\tabularnewline
57624.0 & {\it Swift}/XRT & Interval 2$^\text{d}$ & $\leq$ 106 & $\leq$ 2.2\tabularnewline
\hline
\multicolumn{5}{p{8cm}}{
\textsuperscript{$^\text{a}$}\scriptsize{These values were obtained using a fixed $N_\mathrm{H}$ = 5.7$\times10^{22}$ cm$^{-2}$ for all the spectra.}

\textsuperscript{$\text{b}$}\scriptsize{The unabsorbed luminosity was calculated using a distance of 6.79 kpc.}

\textsuperscript{$\text{c}$}\scriptsize{This interval ranges from 2012 Feb 4 (ObsID: 0031174024) to 2012 Mar 6 (ObsID: 0031174028).}

\textsuperscript{$\text{d}$}\scriptsize{The ObsIDs for this interval are shown in Table \ref{tab_log_obs}.}}
\end{tabular}
\end{table}

We examined the {\it Swift}/BAT and {\it RXTE}/PCA data (when available) of all the other outbursts of SAX J1750. A Sun constraint window occurred during many of its outbursts. We are unable to know how SAX J1750 behaved at these times and so the complete profiles of these outbursts are not known. 
The 2008 outburst shows clear rebrightenings \citep[see Fig. \ref{fig_lc_all}; see also ][]{lowell2012xmm,allen2015spectral}. In the case of the 2001 outburst, the source does not appear to have transitioned fully to quiescence but showed a phase of activity at lower luminosity. However, in both the 2001 and 2008 outburst the main outburst lasted for $\sim$ 200 d and the secondary lower luminosity behaviour for an additional $\sim$ 200 d. 


The peak of the 2015/2016 outburst was followed by a Sun constraint window (Fig. \ref{fig_2008_2015_lc}, bottom panel). It is reasonable to assume that the outburst behaviour followed a gradual decay during this period. After this window the source stayed active for another $\sim$ 200 d. We did not detect any post-outburst rebrightenings after the end of the 2015/2016 outburst during our {\it Swift}/XRT observations. It is possible that we could have missed the rebrightenings because we only have a limited number of {\it Swift}/XRT observations but these rebrightenings seem to last a long time ($\sim$ 200 d as shown by the 2008 outburst). Thus, if the rebrightening behaviour after each outburst has a similar time scale we can exclude the occurrence of rebrightenings after the 2015/2016 outburst. It is unknown what causes the source to sometimes experience a regular outburst and sometimes show rebrightening. Furthermore, considering the full length of the outbursts (main outburst and rebrightenings) the 2001, 2008, and 2015/2016 outbursts are comparable, lasting $\sim$ 300--400 d (see Fig. \ref{fig_lc_all}, \ref{fig_2001_2008_lc}, and \ref{fig_2008_2015_lc}). The 2001 and 2015/2016 outbursts are more similar to each other as compared to the 2008 outburst. They do not experience distinct rebrightenings although their profiles have different shapes -- the main peak of the 2001 outburst is followed by a period of low luminosity before transitioning to quiescence whereas the 2015/2016 outburst does not show this and gradually transitions to quiescence. 

The 2004 and 2011 outburst started $\sim$ 1300 and $\sim$ 1100 d after the 2001 and 2008 outburst, respectively. Both quiescent periods ($\sim$ 1300 and $\sim$ 1100 d) were similar and they were both short outbursts lasting $\sim$ 60 d. Therefore, the source seems to show two different types of outbursts -- long (2001, 2008, and 2015/2016; $\sim$ 400 d) and short outbursts (2004 and 2011; $\sim$ 60 d). The source also exhibited a short outburst ($\sim$ 20 d) in 1997 \citep[see Fig. 2 of][]{natalucci1999new}. About 12 days into this outburst the luminosity decreased (by factor $\sim$ 30) briefly for $\sim$ 5 d before increasing again (by a factor of $\sim$ 4, as observed by the {\it BeppoSAX} Wide Field Cameras [2 -- 30 keV]; \citeauthor{jager1997wide} \citeyear{jager1997wide}). 

\subsection{Quiescent Temperatures and Luminosities}
Here we present the results of the analysis of SAX J1750 in quiescence. All spectra obtained using the different instruments have been fitted using \texttt{XS\textsc{pec}}. We model the equivalent hydrogen column density $N_\mathrm{H}$ with \texttt{tbabs}, using \texttt{VERN} cross-sections  \citep{verner1996atomic} and \texttt{WILM} abundances \citep{wilms2000absorption} for all spectra. W-statistics (background subtracted Cash statistics; \citeauthor{wachter1979parameter} \citeyear{wachter1979parameter}) were used throughout due to the low number of counts per bin for the spectra from the different instruments. The 0.5 -- 10 keV energy range will be considered throughout. All reported fluxes and luminsoities correspond to the unabsorbed ones. The errors are given for the 90 per cent confidence range.

\subsubsection{\textit{XMM-Newton}}
\label{sect_xmm_res}

\begin{figure}
\centering
\includegraphics[scale=0.34]{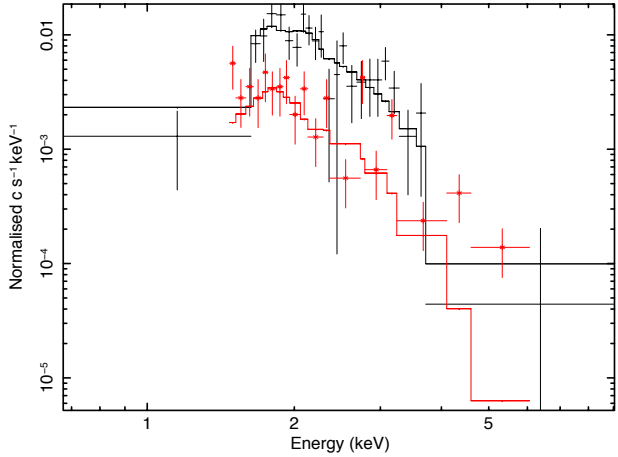}
\caption{The spectrum obtained from the pn instrument during the \textit{XMM-Newton} observation (black +) and from the ACIS-S array during the \textit{Chandra} observation (red $\times$). The best-fit model is shown by the solid lines.}
\label{fig_xmm_spec}
\end{figure}


When we fit the spectra using an absorbed power-law model, the obtained photon index ($\Gamma$ = 6.0$_{-1.1}^{+1.5}$) was very soft (with an $N_\mathrm{H}$ = 7.7$_{-1.8}^{+2.4}\times 10^{22}$ cm$^{-2}$). Due to the softness of the spectra we decided to fit a neutron star atmosphere model (\texttt{nsatmos}; used for NSs with weak magnetic fields; \citeauthor{heinke2006hydrogen} \citeyear{heinke2006hydrogen}) to the spectra. We assumed a neutron star mass and radius of 1.4$M_\odot$ and 10 km, at a distance of 6.79 kpc. The entire surface of the neutron star was assumed to be emitting (i.e., the normalization was set to 1). Initially, the $N_\mathrm{H}$ was left free. The best fit indicated $N_\mathrm{H}$ = (5.7 $\pm$ 0.6) $\times 10^{22}$ cm$^{-2}$ and the observed effective surface temperature was $kT_\text{eff}^\infty$ = 144.9$_{-4.7}^{+4.4}$ eV. We also fit the data with the $N_\text{H}$ fixed to that of the best-fit value, to 5.7 $\times 10^{22}$ cm$^{-2}$, in order to be able to compare the {\it XMM-Newton} results with those obtained from our {\it Chandra} and {\it Swift}/XRT analysis in which we fix the $N_\mathrm{H}$ to this same value (see section \ref{sec_swift_analysis} and \ref{sec_chandra_analysis}). We obtained a $kT_\text{eff}^\infty$ = 144.9$_{-2.5}^{+2.3}$ eV and the flux was $F_\text{X}$ = (1.5 $\pm$ 0.1) $\times 10^{-12}$ erg cm$^{-2}$ s$^{-1}$. The $kT_\text{eff}^\infty$ and luminosity are shown in Table \ref{tab_temp_and_lumin}, and the pn spectrum is shown in Figure \ref{fig_xmm_spec}. All our results are consistent with that of \citet{lowell2012xmm}.\footnote{\citet{lowell2012xmm} found values of $kT_\text{eff}^\infty$ = 148 $\pm$ 4 eV, $N_\mathrm{H}$ = (5.9 $\pm$ 0.5) $\times 10^{22}$ cm$^{-2}$, $F_\mathrm{X}$ = (1.6 $\pm$ 0.2) $\times$ 10$^{-12}$ erg cm$^{-2}$ s$^{-1}$.} Furthermore, \citet{lowell2012xmm} demonstrated that adding a power-law component to the model did not improve the fit to the data.


\subsubsection{{\textit Swift}/XRT}
\label{sec_swift_analysis}
The {\it Swift}/XRT spectra were fitted with \texttt{nsatmos} and the hydrogen column density was fixed to $N_\text{H} = 5.7 \times 10^{22} \text{cm}^{-2}$ (see Section \ref{sect_xmm_res}). For Interval 1 (see Section \ref{sect_obs_xrt}) we found an average count rate of 1.0 $\times 10^{-3}$ c s$^{-1}$. The observed effective temperature of the source in Interval 1 was $kT_{\text{eff}}^{\infty} = 130.6_{-12.5}^{+10.6}$ eV and the flux was $F_\text{X} = 9.7_{-4.7}^{+6.2} \times 10^{-13}$ erg cm$^{-2}$ s$^{-1}$ (see also Table \ref{tab_temp_and_lumin}). Our results for this interval are consistent with those of \citet{wijnands2013low}\footnote{\citet{wijnands2013low} found values of $kT_\text{eff}^\infty$ = 150 $\pm$ 20 eV, for $N_\mathrm{H}$ fixed to 6 $\times 10^{22}$ cm$^{-2}$. The $F_\mathrm{X}$ was (1.6 $\pm$ 0.7) $\times$ 10$^{-12}$ erg cm$^{-2}$ s$^{-1}$.}. 

In the post 2015/2016 outburst stacked observations (Interval 2; see Table \ref{tab_log_obs}) the source was not detected. We used the prescription outlined by \citet{gehrels1986confidence} to calculate the count rate upper limit for a 95 per cent confidence level. The obtained upper limit on the count rate was $\leq 0.4 \times 10^{-3}$ c s$^{-1}$. To get an upper limit on the observed effective surface temperature we simulated a spectrum using \texttt{fakeit} in \texttt{XS\textsc{pec}}. The simulation used the \texttt{arf} of the stacked observations in Interval 2 and the appropriate response matrix file (swxpc0to12s6$\_$20130101v014.rmf). For the simulated spectrum, we employed an \texttt{nsatmos} model with parameters consistent with those we have used so far, such as those for mass, radius, normalization, distance, and $N_\mathrm{H}$. We then adjusted the \texttt{nsatmos} temperature such that the model predicted count rate within \texttt{XS\textsc{pec}} matched our calculated count rate upper limits. Using this method, the upper limit on the temperature was determined to be $\leq$ 106 eV and the upper limit on the flux was $F_\text{X} \leq 4.0 \times 10^{-13}$ erg cm$^{-2}$ s$^{-1}$ (see Table \ref{tab_temp_and_lumin} for the luminosity upper limit).

\subsubsection{\textit{Chandra}}
\label{sec_chandra_analysis}
We fit the {\it Chandra} spectrum with an \texttt{nsatmos} model and obtained an observed effective temperature of $kT_{\text{eff}}^{\infty} = 129.9_{-5.7}^{+5.3}$ eV with a column density value (consistent with that from our {\it XMM-Newton} analysis) of $N_\mathrm{H}$ = (6.3 $\pm$ 0.8) $\times 10^{22}$ cm$^{-2}$. When fixing the column density to the value determined using {\it XMM-Newton}, we obtained a $kT_{\text{eff}}^{\infty} = 126.2_{-3.3}^{+3.2}$ eV and a flux of $F_\text{X} = 7.6_{-1.2}^{+1.3} \times 10^{-13}$ erg cm$^{-2}$ s$^{-1}$. The inferred luminosity was $L_\text{X}$ = (4.2 $\pm$ 0.7) $\times 10^{33}$ erg s$^{-1}$ (D/6.79 kpc)$^2$ (see also Table \ref{tab_temp_and_lumin}). The spectrum is shown in Figure \ref{fig_xmm_spec}. We added a power-law component to the model to determine if it was necessary and refitted the spectrum. We found that this additional component was not required to improve the fit. We determined the upper limit on the flux contribution from this component to be $\sim$ 10 per cent (assuming a photon index of 1.5).

%

\section{Discussion}
We studied the end of outburst behaviour of SAX J1750 after several outbursts  (see Section \ref{sect_disc_outburst}) as well as the quiescent properties of the source (see Section \ref{sect_disc_quiescence}).

\subsection{Outburst variability}
\label{sect_disc_outburst}
Several dwarf novae and transient LMXBs (both neutron star and black hole ones) show post-outburst rebrightenings. The NS LMXB SAX J1808.4$-$3658 likely showed such rebrightenings after all its outbursts \citep[][]{patruno2016reflares}. SAX J1750 previously showed a post-outburst rebrightening after its 2008 outburst \citep{lowell2012xmm,allen2015spectral}. We requested {\it Swift}/XRT observations of SAX J1750 at the end of its 2015/2016 outburst to determine if post-outburst rebrightenings are also a recurrent feature in SAX J1750, however, we found that it is not the case for this source.

The behaviour of SAX J1750 during different outbursts differs significantly from each other. Some outburst are relatively short ($\leq$ 1 month) while other outburst last  $>$ 1 yr.  In addition, at least one long outburst (2008) showed an episode of rebrightening while for the last long outburst (2015/2016) of the source such rebrightenings were not observed. The reason(s) for this different outburst behaviour is not understand but any model explaining the outburst behaviour of this source has to take these outburst variations into account. We note that many other X-ray transients show quite a variety of outburst profiles so SAX J1750 is not unusual in its behaviour \citep[e.g., neutron star sources such as Aql X$-$1 and 4U 1608$-$52, and black hole source such as 4U 1630$-$47 and GX 339$-$4; ][]{lei2014properties,capitanio2015missing,clavel2016systematic,waterhouse2016constraining}. Furthermore, \citet{wijnands2013low} reported a flare during their {\it Swift}/XRT observations after the 2011 outburst, increasing the complexity of the accretion behaviour of this source. Similar flares have been seen for other sources as well (e.g., Aql X-1, \citeauthor{coti2014year} \citeyear{coti2014year}; see \citeauthor{wijnands2013low} \citeyear{wijnands2013low} for an in depth discussion). The flaring behaviour also needs to be accounted for in a model attempting to explain the accretion variability in SAX J1750.

\subsection{SAX J1750 in quiescence}
\label{sect_disc_quiescence}
SAX J1750 was observed in quiescence after 3 outbursts -- after the 2008 outburst using {\it XMM-Newton}, after the 2011 outburst using {\it Swift}/XRT and {\it Chandra}, and after the 2015/2016 outburst once again using {\it Swift}/XRT. These are all the known outbursts of this source since 2008. However, if accretion activity occurred in one of the Sun constraint windows or was not at a high enough level to be picked up by BAT then it could have been missed and might complicate the discussion. For our discussion we assume that no undetected activity occurred.
 
Our quiescent data can be well fit by a thermal component model, without the need of a power-law component. This thermal emission likely comes from the neutron star surface caused by a cooling emission from a hot neutron star or low level accretion of matter onto the surface. The latter indeed could potentially produce an entirely thermal spectrum \citep[see e.g.,][]{zampieri1995x} and the quiescent variations we see can then be explained by fluctuations in the residual accretion rate. However, recently it has been suggested that low level accretion onto a neutron star in quiescence not only produces a soft component but also a hard component \citep[e.g.,][]{chakrabarty2014hard,d2015radiative,wijnands2015low} that is roughly equal in strength in the 0.5 -- 10 keV range compared to the soft component \citep[see discussion in][]{wijnands2015low}. If true, the absence of the power-law component in our spectra would indicate that we do not see low-level accretion but indeed cooling emission from the surface. In the remainder of the discussion we will consider our observations in the context of this assumption.

\subsubsection{Do we see crust heating and cooling during our observations?}
The {\it XMM-Newton} observation taken $\sim$ 400 d after the end of the 2008 outburst (and rebrightenings) indicated an observed effective neutron star temperature $kT_{\text{eff}}^{\infty}$ of $\sim$ 145 $\pm$ 2.5 eV, consistent with the results of \citet{lowell2012xmm}. They proposed that at the time of the observation the crust and core were in equilibrium. From the very high surface temperature, they also inferred that SAX J1750 may harbour the hottest known NS in a LMXB.
 \citet{wijnands2013low} also observed SAX J1750 in quiescence, $\sim$ 350 d after the end of the small 2011 outburst, using {\it Swift}/XRT. We analysed the pre-flare data to determine the quiescent crust temperature of SAX J1750. We found a $kT_{\text{eff}}^{\infty}$ of $\sim$ 131 $\pm$ 12 eV, consistent with the results of \cite{wijnands2013low} and consistent with the value seen by \cite{lowell2012xmm}. This further supported the hypothesis that SAX J1750 may host the hottest known NS in a LMXB as after a different outburst the source was still at a similar temperature to that after the 2008 outburst (although we note that the very large error bars on these post 2011 data are not very constraining). 
 
At the time of publication by \citet{lowell2012xmm} it was generally assumed that short outbursts (of only a few months) could not heat a NS crust significantly out of equilibrium with the core. And therefore, this study assumed crust-core equilibrium when interpreting their results. Since then several sources have shown that in fact such short outbursts can still heat the crust significantly out of equilibrium with the core \citep{degenaar2013continued,degenaar2015neutron,waterhouse2016constraining,parikh2017potential}. The dominant source of this heating is postulated to be located at a shallow depth in the crust. This is different from the deep crustal heating reactions which will still contribute to crustal heating during such short outbursts but not dominate it. The origin of this shallow heat source is unknown but has been observed for many sources. Typically it contributes up to $\sim$ 2 MeV per accreted nucleon of heat \citep{brown2009mapping,degenaar2011evidence,degenaar2014probing}. Therefore, it is possible that for the {\it XMM-Newton} and post 2011 outburst {\it Swift}/XRT observations we still observed a heated crust rather than the crust in equilibrium with the core. If so, this temperature may not be directly related to the core temperature (see discussion below) and we cannot conclude that SAX J1750 is the hottest known NS LMXB. We further discuss the possibility of SAX J1750 hosting the hottest NS in Section \ref{sec_hottest_ns}.
 
To further study the quiescent behaviour of this source we searched the various satellite archives. We found a {\it Chandra} observation of the source taken $\sim$ 800 d after the end of the short 2011 outburst and we found a $kT_{\text{eff}}^{\infty}$ of $\sim$ 126 $\pm$ 3 eV. If no undetected accretion activity in between the Interval 1 {\it Swift}/XRT observations and the {\it Chandra} observation, this {\it Chandra} observation was taken further into quiescence after the end of the 2011 outburst compared to the Interval 1 {\it Swift}/XRT observations and could be indicative of a cooling crust.
 However, due to the large error bars on these {\it Swift}/XRT data we are unable to get an exclusive interpretation. The crust at the time of the {\it Chandra} observation could be at the same level, or lower, or even higher than at the time of the {\it Swift}/XRT observation. Therefore, we do not discuss the {\it Swift}/XRT data from Interval 1 further in context with the {\it XMM-Newton} and {\it Chandra} data because of its large error bars. 
 
The {\it XMM-Newton} observation was taken $\sim$ 400 d after a long, bright 2008 outburst and may show a heated NS crust. Unfortunately, no follow up observations are present in the same quiescent period to check for crust cooling. The {\it Chandra} observation was taken further into quiescence, $\sim$ 800 d after the shorter 2011 outburst. The brightness of this outburst is not known due to limitations introduced by the Sun constraint window. The {\it Chandra} observation indicates a lower effective observed temperature compared to the {\it XMM-Newton} observation potentially indicating a cooling crust (although the short 2011 outburst may have reheated the crust to some degree).

SAX J1750 experienced another relatively long, bright outburst in 2015/2016. When using {\it Swift}/XRT to observe SAX J1750 after this outburst we did not detect the source. These observations were done very close after the estimated end of outburst ($\sim$ 50 d after the end). From these observations we obtained an upper limit on the observed effective temperature of $\leq$ 106 eV. This upper limit is significantly lower than previously obtained quiescent temperatures. However, it is not trivial to explain this low quiescent temperature.
 One of the explanations of such a relatively low temperature upper limit so soon after the end of outburst would be that the crust of the neutron star was not heated during the preceding accretion outburst. Similar behaviour has been observed for MAXI J0556$-$332, which experienced three outbursts. Modelling the quiescent evolution for this source indicated that the outbursts (having different profiles) need different magnitudes of shallow heating to explain the post-outburst cooling trends \citep{homan2014strongly,deibel2015strong,parikh2017cooling}. In particular, the cooling curve after outburst two of MAXI J0556$-$332 can be modelled without any shallow heating. A similar effect might be at work in SAX J1750, although it remains unclear what causes this difference in the amount of heat generated by the shallow heating process during the different outbursts, in particular because both the 2008 and the 2015/2016 had similar properties (i.e., peak brightness, outburst durations and profiles) so one would expect similar heating during those outbursts. However, since this process is still far from understood, currently we cannot exclude this possibility.
 
A change in the envelope composition is another possibility to explain the relatively low temperature limit observed for SAX J1750 so soon after the end of the 2015/2016 outburst. The inferred neutron star crust temperature seen by an observer depends on the chemical composition of its envelope \citep{potekhin1997internal}. We refer to the outer $\sim$ 100 m of the neutron star as the envelope \citep[where $\rho \lesssim$ 10$^{8}$ g cm$^{-3}$,][]{potekhin2015neutron}. The envelope acts as a blanket and has a strong thermal gradient. \citet{brown2002variability} showed that the observed effective surface temperature seen by the observer depends on the amount of hydrogen and helium left over in the envelope after the nuclear burning that occurred during accretion. If the envelope consists mostly of light element materials the observed effective temperature is higher than if the envelope has mostly heavy elements. It has been shown that a change in the envelope composition from light elements to heavy elements can cause the luminosity to drop by a factor of a few \citep{brown2002variability,page2004minimal,han2017cooling}. Therefore, conceivably, our results could be explained by a more light element composition in the envelope after the 2008 outburst compared to after the 2015/2016 outburst. However, detailed modelling needs to be performed, also taking into account variations in the properties (i.e., strength and depth) of the shallow heating, to determine if crust heating and cooling can explain our quiescent observations. 

Our observations of SAX J1750 can possibly be explained by some combination of cooling of an accretion-heated crust accompanied by a changing envelope composition. However, it may be that this model has to be fine-tuned to be able to explain the observations. Detailed modelling should be performed to determine if the assumptions that need to be made in this case are realistic.
  

\subsubsection{The hottest NS in SAX J1750}
\label{sec_hottest_ns}
The {\it Chandra} and the 2016 {\it Swift}/XRT observations show the NS in SAX J1750 to be at temperatures much lower than $\sim$ 145 eV, based on which \citet{lowell2012xmm} proposed it to be the hottest known NS LMXB. Can it still be the hottest known NS LMXB? If the various different temperatures can be explained only by a change in the chemical composition of the envelope it could still be the hottest known NS LMXB. Although it seems unlikely that the envelope composition after the three outbursts changes in a manner such that we always observe a decreasing temperature. Alternatively, if these different temperatures are the result of a cooling NS crust SAX J1750 will not host the hottest NS in a LMXB.

In addition, one must also consider the effect of the envelope composition on the other quiescent NS LMXB observations. Ideally one has to compare NS LMXBs in quiescence which have the same envelope composition. This is of course not possible because it is difficult to determine the composition, introducing another uncertainty in the models \citep[see][]{han2017cooling}. These other NS LMXBs may be hotter than SAX J1750 but if their envelopes have heavier elements the observed effective temperature would be lower than if they had a pure light element envelope. We must take this uncertainty into account when discussing the possibility of the hottest NS LMXB and at all other times when conclusions are being drawn based the the absolute value of the observed effective temperature. This will also affect the conclusions that can be inferred when creating  the quiescent luminosity versus the long-term average accretion rate diagram commonly used in quiescent NS LMXBs studies to infer NS core properties (e.g., \citeauthor{heinke2010discovery} \citeyear{heinke2010discovery}; see also \citeauthor{han2017cooling} \citeyear{han2017cooling} for such a comparision).


\section*{Acknowledgements}

AP and RW are supported by a NWO Top Grant, Module 1, awarded to RW. We acknowledge the use of public data from the {\it Swift} data archive. This research has made use of the {\it Swift}/BAT transient monitor results provided by the {\it Swift}/BAT team, as well as NASA's Astrophysics Data System.



\bibliographystyle{mnras}


 \newcommand{\noop}[1]{}


\bsp	
\label{lastpage}
\end{document}